\documentstyle[emulateapj,epsf]{article}

\newcommand{\bvec}[1]{\mbox{\boldmath$#1$}}

\begin{document}
\title{Robustly Unstable Eigenmodes\\
of the Magnetoshearing Instability
in Accretion Disk}
\author{K. Noguchi\footnotemark[1] and T. Tajima}
\affil{Department of Physics and Institute for Fusion Studies, Univerisity of Texas at Austin, Austin, TX 78712}

\and

\author{R. Matsumoto}
\affil{Department of Physics, Chiba University, 1-33 Yayoi-cho, Inage-ku, Chiba 263, Japan}
\footnotetext[1]{knoguchi@mail.utexas.edu}
\begin{abstract}
The stability of nonaxisymmetric perturbations in differentially rotating
astrophysical accretion disks is analyzed by fully incorporating the 
properties of shear flows.
We verify the presence of discrete unstable
eigenmodes with complex and pure imaginary eigenvalues, without any 
artificial disk edge boundaries, unlike Ogilvie \& Pringle(1996)'s claim. 
By developing the mathematical theory of a non-self-adjoint system, 
we investigate the nonlocal behavior of eigenmodes in the vicinity 
of Alfv\'en singularities at $\omega_D=\pm\omega_A$, where $\omega_D$ is the Doppler-shifted
wave frequency and $\omega_A=k_\parallel v_A$ is the Alfv\'en frequency. 
The structure of the spectrum of discrete eigenmodes is discussed
and the magnetic field and wavenumber dependence of the growth rate are obtained.
Exponentially growing modes are present even in a region where the local 
dispersion relation theory claims to have stable eigenvalues.
The velocity field created by an eigenmode is obtained, which explains 
the anomalous
angular momentum transport in the nonlinear stage of this stability. 
\end{abstract}
\keywords{accretion, accretion disks --- instabilities --- MHD --- plasmas}
\nopagebreak
\section{Introduction}
Over the last several years, the presence of magnetic fields in a differentially rotating
plasma has been proposed as a possible mechanism of accretion disk
turbulence and its associated large anomalous angular momentum transport
inside the disk. The presence of magnetic fields in 
a shear rotating gas cylinder makes the gas unstable against axisymmetric perturbations(\cite{vel59}; \cite{cha61}).
The normal mode analysis(\cite{kum94}) of this local magnetoshearing instability
showed the existence of unstable axisymmetric eigenmodes. The presence of this robust instability
was re-recognized (\cite{bal91}) and confirmed by nonlinear ideal MHD simulations(\cite{haw91}, 1992; \cite{haw95}). The observational features of astrophysical accretion disks point to the need for large viscosity much beyond the one collisional mechanisms can yield. This robust instability has been invoked (\cite{haw92}) as a most promising candidate mechanism for the viscosity puzzle(e.g., \cite{taj97}).

Numerical investigation of nonaxisymmetric magnetoshearing modes has been carried out by adopting the shearing coordinates(e.g., \cite{bal92}).
Matsumoto \& Tajima (1995) analyzed nonaxisymmetric nonlocal eigenmodes 
which are sandwiched by two Alfv\'en singularities around the corotational point
and are not influenced by the disk edge boundaries and grow exponentially in time.
These modes are distinct from the modes discussed by Ogilvie \& Pringle (1996), which are nonaxisymmetric modes contained within cylindrical boundaries and which depend strongly on the boundary conditions. 

In this paper, we concentrate our focus on the validity of the linear analysis of nonaxisymmetric eigenmodes 
questioned by Ogilvie \& Pringle(1996). The analysis is performed in the frame rotating with the local angular velocity, which is adopted in nonaxisym\-met\-ric mode analysis(\cite{ogi96}; \cite{mat95}), since eigenmodes evolve exponentially in time.
The resolution of this question is important in theory of accretion disk Unless this magnetoshearing instability is a robust mode unaffected by the boundary conditions, the long search for candidate mechanisms of anomalous viscosity of
accretion disks needs to be reopened. The criticism of Ogilvie \& Pringle (1996) is interesting because it reflects the difficulty and the extreme mathematical and physical subtlety involved in the nature of this mode around the Alfv\'en
singularity. The problem is marred by the non-self-adjointness of the differential equation that describes eigenmodes, arising from the presence of shear flows.
We know of no systematic mathematical theory on non-self-adjoint differential equations. Thus it takes us a development of mathematical and physical theory of such a system in order to understand the argument by Ogilvie \& Pringle and to respond aptly to it. In the end such analysis has been developed and,
assuringly, we find that the magnetoshearing instability is both robust and insensitive to the boundary condition, as we thought originally.
 
In \S 2, we derive the wave equation in a differentially rotating magnetized disk based on the
analysis of Matsumoto \& Tajima (1995), and basic properties of the wave equation are discussed. We show that discrete nonaxisymmetric eigenmodes exist, which are buffeted by the pair of Alfv\'en singularities where the Doppler-shifted wave frequency equals the Alfv\'en frequency. Our analysis finds that the eigenmodes oscillate indefinitely in the vicinity of the Alfv\'en singular points when the eigenvalue is real, whereas the eigenmodes are regular when the eigenvalue is complex.
Numerical calculation of eigenmodes with the Alfv\'en frequency and wavenumber dependence is discussed in \S 3. We compare the results with the local dispersion relation, and show these eigenmodes are discrete. Astrophysical implications and conclusions are discussed in \S 4.  

\section{Analytical Properties of Non-Self-Adjoint Equation near Alfv\'en Singularity}
We consider the MHD stability of magnetoshearing modes in the co-rotating frame of the fluid. The basic ideal MHD equations in the frame rotating with angular velocity $\Omega$ are
\begin{eqnarray}
\left(\frac{\partial}{\partial t}+(\bvec{v}\cdot\nabla)\right)\bvec{v}&=&-\frac1{\rho}\nabla P+\frac{\nabla\times\bvec{B}\times\bvec{B}}{4\pi\rho}+\bvec{g}\nonumber\\
&+&2\bvec{v}\times\bvec{\Omega}+(\bvec{\Omega}\times\bvec{r})\times\bvec{\Omega},
\end{eqnarray}
\begin{equation}
\frac{\partial \bvec{B}}{\partial t}\!=\!\nabla\times(\bvec{v}\times\bvec{B}),
\end{equation}
where $\bvec{g}$ is the gravitational
acceleration and $\bvec{r}$ is the position vector. We assume incompressibility for simplicity,
\begin{equation}
\nabla\cdot\bvec{v}=0.\label{cont}
\end{equation}
We also ignore self gravity, which is not essential for the magnetoshearing
instability.

We use the local Cartesian coordinates ($x, y, z$) in the rotating frame where the $x$-axis is in the 
radial direction, the $y$-axis in the azimuthal direction, and the $z$-axis parallel to $\Omega$. 
The uniform velocity shear $v_y=-(3\Omega/2)x$ is assumed for the Keplerian disk, where $x=0$ is the local co-rotating radial position. The wave equation is
derived by linearizing the basic equations around the equilibrium state and assuming solution of the
form $\tilde{\phi}(x,t)\mbox{exp}[\mbox{i}(k_yy+k_zz)]$. In the unperturbed state, the density, pressure and
magnetic field are assumed to be uniform. The assumption of $v_x=v_z=B_x=0$ in the unperturbed state yields the unperturbed momentum equation,
\begin{equation}
\bvec{g}+2\bvec{v}_0\times\bvec{\Omega}+(\bvec{\Omega}\times\bvec{r})\times\bvec{\Omega}=0.
\end{equation}

Next, the Laplace transform of the perturbation, $\bar{\phi}(x,\omega)$, is employed,
\begin{equation}
\bar{\phi}(x,\omega)=\int^{\infty}_0dt{}\tilde{\phi}e^{\mbox{i}\omega t}.
\end{equation}
Substitution of the Laplace transformed momentum and induction equations into the continuity equation
yields (see \cite{mat95} for detail) the initial value equation
\begin{eqnarray}
\lefteqn{\frac{d^2\bar{v}_x}{d{}x^2}+\frac{3\Omega\omega_A^2k_y}{\omega_D(\omega_D^2-\omega_A^2)}\frac{d\bar{v}_x}{d{}x}+\Bigg[-(k_y^2+k_z^2)}\nonumber\\
&&\!\!\!\!\!\!\!\!\!\!\!\left.-\frac{9\Omega^2k_y^2\omega_A^2}{2\omega_D^2(\omega_D^2\!\!-\!\omega_A^2\!)}+\Omega^2k_z^2\frac{\omega_D^2\!+3\omega_A^2}{(\omega_D^2\!\!-\!\omega_A^2\!)^2}\!\right]\!\!\bar{v}_x\!=\!\Gamma(x,\omega),\label{eq1}
\end{eqnarray}
where $\omega_D$ is the Doppler-shifted frequency, 
\begin{equation}
\omega_D=\omega+\frac{3}{2}\Omega k_y x,\label{omd}
\end{equation}
and $\omega_A$ is the Alfv\'en frequency,
\begin{equation}
\omega_A^2=\frac{(\bvec{k}\cdot\bvec{B})^2}{4\pi\rho}=k_\parallel^2v_A^2.\label{oma}
\end{equation}
The initial condition enters through the source function $\Gamma(x,\omega)$.

The wave equation is derived by expressing the homogeneous part of equation (\ref{eq1}) in 
terms of the normalized radial coordinate
\begin{equation}
\xi=\frac{3\Omega k_y}{2\omega_A}x
\end{equation}
as
\begin{eqnarray}
\lefteqn{\frac{d^2\bar{v}_x}{d\xi^2}+
\frac{2\omega_A^3}{\omega_D(\omega_D^2\!-\!\omega_A^2)}\frac{d\bar{v}_x}{d\xi}+\!\!\Bigg[\!\!-\frac{4}{9}\left(1+\frac{1}{q}\right)\!\!\left(\displaystyle{\frac{\omega_A}{\Omega}}\right)^{\!2}}\nonumber\\
&&
\left.-\frac{2\omega_A^4}{\omega_D^2(\omega_D^2-\omega_A^2)}
+\frac{4}{9}\frac{\omega_A^2}{q}\frac{\omega_D^2+3\omega_A^2}{(\omega_D^2-\omega_A^2)^2}\right]\bar{v}_x\nonumber\\
&&\equiv D(\omega,\xi)\bar{v}_x=0,\label{eq2}
\end{eqnarray}
where the ratio of the squares of the azimuthal and vertical wavenumber is defined as
\begin{equation}
q=\frac{k_y^2}{k_z^2}.
\end{equation}

Unstable eigenmodes may exist when the solution satisfies the boundary condition
\begin{equation}
\lim_{|\xi|\to\infty}\bar{v}_x=0\label{bound}
\end{equation}
in the upper half of the complex $\omega$-plane reference. This boundary condition makes our eigenmodes distinct from the modes found by Ogilvie \& Pringle (1996), which are confined in rigid cylindrical boundaries and strongly dependent on the boundary condition. In order to have an overall angular momentum transport across the entire disk, it is imperative to have unstable modes within the disk, not just on the boundaries of the disk. To investigate the interior of the accretion disk,
eigenmodes should not depend on the edge boundary conditions.
The eigenmodes which arise from finite boundaries may contribute to the angular momentum only near the edge of the disk. Our mode, which is not affected by the edge, grows whenever the eigenfunction is located between two Doppler-shifted Alfv\'en points.
Since the positions of those points are determined in co-rotating frame, the origin of the frame can be anywhere in the disk. The boundary
condition then assures that the momentum transport resulted from the superposition of the growing eigenmodes which can occur throughout the accretion disk.

The proper boundary condition interior of the disk can be easily examined by inspecting the aysmptotic form of the basic differential equation (\ref{eq1}) of the system. This indicates that the leading radial dependence of eq. (\ref{eq1}) leads to the exponential decay away from the co-rotation point and toward the Alfv\'en singularity. This mathematics is, of course, most reasonable and physical as well, because the instability energy is provided within the two Alfv\'en singular layers to the mode, which dissipates the energy at(or near) the singularities.

Since the wave equation (\ref{eq2}) is not self-adjoint due to the existence of the flow shear, the square of the eigenvalue $\omega^2$ is not guaranteed to be real, which any self-adjoint system always satisfies. The fact that the eigenvalue is not pure real or imaginary but in general complex prevents us from applying the Strum-Liouville theory to this system. Note that if $k_y=0$, equation (\ref{eq1}) is reduced to self-adjoint, 
\begin{equation}
\frac{d^2\bar{v}_x}{d x^2}+k_z^2\left[-1+\Omega^2\frac{\omega^2+3\omega_A^2}{(\omega^2-\omega_A^2)^2}\right]\bar{v}_x=0,
\end{equation}
and its eigenvalue constitutes the Alfv\'en continuum. The analysis of this mode has been done by Chandrasekhar(1961). We now assume $k_y\neq0$.

To the best of our knowledge, no theory for non-self-adjoint operators exists, and we have to investigate the properties of non-self-adjoint systems in general.
We find that the eigenmodes possess certain symmetry properties, which are originated in the character of the differential operator and radial symmetry.
Since the differential operator $D(\omega, \xi)$ is invariant under the operation $(\omega,\xi)\to(-\omega,-\xi)$ because of radial symmetry,
the eigenfunction $\bar{v}_x(\omega,\xi)$ is also the eigenfunction of $D(-\omega, -\xi)$,
\begin{equation}
D(-\omega, -\xi)\bar{v}_x(\omega,\xi)=0.\label{eq3}
\end{equation}
Note that the operation $\xi\to-\xi$ is the same as changing the direction of the rotation $\Omega\to-\Omega$. Taking conjugate of the equation (\ref{eq3}) and denoting $-\xi$ to $\xi$ yields
\begin{eqnarray}
\lefteqn{D(-\omega^*,\xi)\bar{v}_x^*(\omega,-\xi)}\nonumber\\
&&=D(-\omega^*,\xi)\bar{v}_x(-\omega^*,\xi)=0,\label{eq4}
\end{eqnarray}
where ${\phi}^*$ means the complex conjugate of $\phi$. Thus, if $\omega$ is an unstable eigenvalue of the equation (\ref{eq2}), $-\omega^*$ is another unstable eigenvalue whose eigenfunction $\bar{v}_x(-\omega^*,\xi)$ satisfies the relation (\ref{eq4}).
Especially, when $\omega$ is pure imaginary, i.e., $-\omega^*=\omega$, the real part of eigenfunction is symmetric and the imaginary part antisymmetric with respect to $\xi=0$,
\begin{equation}
\bar{v}_x^*(\omega,-\xi)=\bar{v}_x(\omega,\xi).\label{sym1}
\end{equation}
These properties of the non-self-adjoint operator indicate that this system has complex eigenvalues in general, which differs from a self-adjoint system, and one unstable eigenvalue has another unstable and two stable companions. 
This symmetry of non-self-adjoint system and the comparison between self-adjoint and non-self-adjoint eigenvalues in complex-$\omega$ plane is shown in Figure 1.
\begin{figure*}[htb]
\epsfxsize=3 truein \centerline{\epsffile{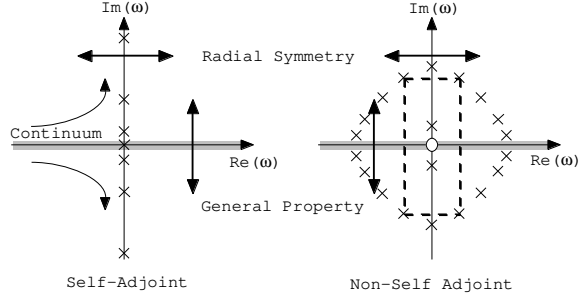}}
\caption{Symmetry of a nonself-adjoint system and the comparison between self-adjoint
and nonself-adjoint eigenvalues in complex-$\omega$ plane. Discrete modes are denoted by crosses. The eigenvalue of the self-adjoint operator should be purely real or
imaginary, and the eigenfunctions are predictable by Strum-Liouville theory. However,
there is no such restriction for the eigenvalue of the nonself-adjoint operator, 
and $\omega,\omega^*,-\omega$ and $-\omega^*$ make a group of solutions.
}
\label{fig1}
\end{figure*}
 
Next, we look for the solution of the wave equation (\ref{eq2}).
The boundary conditions for these ideal MHD modes we are interested in are
that the eigenmode resides and is differentiable around the corotational
point and sandwiched by a pair of the Alfv\'en singularities, and decays
toward $\xi\to\pm\infty$. Note that these physical boundary conditions
preclude the usual Kelvin-Helmholtz(K-H) instability eigenmodes (\cite{taj91};
\cite{nog98}). The boundary conditions generic to the
K-H mode is to match the dissipative structure at the corotational point.
The properties of these eigenmodes may be investigated by analyzing the
solution around the spatial coordinates of interest by using the Frobenius expansion (\cite{mor53}),
in particular around the Alfv\'en singular points. It is of particular
significance to examine analytical properties of the Frobenius expression
around the Alfv\'en singularities in order to examine the question 
leveled by Ogilvie \& Pringle (1996) against the eigenmodes obtained by Matsumoto
\& Tajima(1995). An appropriate, compatible numerical method for these
eigenmodes is, therefore, the shooting method that starts from an exponentially
decaying functional form at $\xi=\pm\infty$ and shoots toward the corotating
point where two sides of the function should smoothly (differentiably)
match.(This will be closely examined in the next section.)

We expand $\bar{v_x}$ in equation (\ref{eq2}) using,
\begin{equation}
\bar{v}_x=\sum_{n=0}^{\infty} a_n(\xi-\xi_{A\pm})^{n+s}
\end{equation}
in the vicinity of the Alfv\'en singularities $\omega_D=\pm\omega_A$ or $\xi=\xi_{A\pm}$, 
where $\xi_{A\pm}$ is defined by
\begin{eqnarray}
\xi_{A\pm}&=&\pm1-\frac{\omega}{\omega_A}\nonumber\\
&=&\pm1-\omega'_r-\mbox{i}\omega'_i,
\end{eqnarray}
and $\omega'_r(\omega'_i)$ is the real(imaginary) part of $\omega/\omega_A$.
Applying the Frobenius method in the vicinity of the regular singular points $\xi=\xi_{A\pm}$, the indicial equation for the exponent $s$ for Eq. (\ref{eq2}) is given by
\begin{equation}
s^2+\frac{\omega'_i(\omega'_i\pm3\mbox{i})}{2\mp3\mbox{i}\omega'_i-{\omega'_i}^2}s
+\frac{4}{9q}\left[\frac{4\mp2\mbox{i}\omega'_i-{\omega'_i}^2}{(2\mp\mbox{i}\omega'_i)^2}\right]=0.
\end{equation}
Note that the eigenfunction is irregular and oscillates indefinitely at $\xi=\xi_{A\pm}$ whenever $s$ has imaginary component, provided Re($s$) is nonpositive integer,
\begin{equation}
\bar{v}_x=(\xi-\xi_{A\pm})^{\mbox{\tiny Re}(s)}\exp\left[\mbox{Im}(s)\log|\xi-\xi_{A\pm}|\right].\label{osc}
\end{equation}
The singular points are on the real axis if and only if the eigenvalue $\omega$ is real.
In general, the indices are given by
\begin{eqnarray}
\lefteqn{s=\frac{\omega'_i(\omega'_i\pm^*3\mbox{i})}{2(2\mp^*\mbox{i}\omega'_i)(1\mp^*\mbox{i}\omega'_i)}\times\Bigg[-1}\nonumber\\
&&\left.\pm\sqrt{1-\frac{16}{9q}\frac{(4\mp^*2\mbox{i}\omega'_i-{\omega'_i}^2)(1\mp^*\mbox{i}\omega'_i)^2}{{\omega'_i}^2(\omega'_i\pm^*3\mbox{i})^2}}\right],\label{sorg}
\end{eqnarray}
where the sign $\pm^*$ indicates that we take upper sign when $\xi=\xi_{A+}$, and lower sign when $\xi=\xi_{A-}$, respectively. Since $s$ is complex, the solution is not analytic at the Alfv\'en singular points.

Now, we investigate some special cases. First, when the eigenvalue is pure real, i.e., $\omega'_i=0$, the indices are purely imaginary,
\begin{equation}
s=\pm\frac{2}{3\sqrt{q}}\mbox{i},\label{real}
\end{equation} 
the eigenfunction rapidly oscillates and is indefinite. Note that the boundary condition in this case becomes special in that we need to shoot outward from $\xi=\xi_c$. We can no longer shoot from the outer to the inner region. 
As the eigenmodes oscillate indefinitely around the Alfv\'en singularity, we find that the eigenvalue becomes continuous in this pure real case to form the Alfv\'en continuum. 
Second, when $\omega'_i\ll1$, which is similar to the first case but now the singular points are not on the real axis, the solutions are 
\begin{equation}
s=\pm\left[\frac{2}{3\sqrt{q}}\mbox{i}\mp^*\frac{3}{32}\sqrt{q}\omega'_i\right],\label{s}
\end{equation}
whose solution oscillates limited times in the vicinity of the singular points on the real axis, but not indefinitely, and the solution is regular on the real axis.
Third, when the perturbation is nearly toroidal ($k_y\gg k_z, q\gg1$), as in the accretion disks far from their source, and the eigenvalue is complex, $s$ is
given by
\begin{eqnarray}
\lefteqn{s=\frac{\omega'_i(\omega'_i\pm^*3\mbox{i})}{2(2\mp^*\mbox{i}\omega'_i)(1\mp^*\mbox{i}\omega'_i)}\times\Bigg[-1}\nonumber\\
&&\!\!\!\!\!\left.\pm\left(1-\frac{8}{9q}\frac{(4\mp^*2\mbox{i}\omega'_i-{\omega'_i}^2)(1\mp^*\mbox{i}\omega'_i)^2}{{\omega'_i}^2(\omega'_i\pm^*3\mbox{i})^2}\right)\right]\!.\label{qinf}
\end{eqnarray}
When the perturbation is pure toroidal, the second term in the square brackets in equation (\ref{qinf}) vanishes and $s$ is given by
\begin{equation}
s=0,\quad-\frac{\omega'_i(\omega'_i\pm^*3\mbox{i})}{(2\mp^*\mbox{i}\omega'_i
)(1\mp^*\mbox{i}\omega'_i)}.
\end{equation}
The eigenfunction corresponding to $s=0$ is regular, and the latter one diverges logarithmically at the singular point, when $\omega$ is real.
Finally, when the perturbation is nearly poloidal($k_y\ll k_z,q\ll1$), as may occur in accretion disks close to their source, we find $s$ as
\begin{equation}
s=\pm\frac{2\mbox{i}}{3\sqrt{q}}\frac{\sqrt{4\mp^*2\mbox{i}\omega'_i-{\omega'_i}^2}}{2\mp^*\mbox{i}\omega'_i},
\end{equation}
which reduces to the roots of the first case when the eigenvalue is pure real.

The exponent $s$ at the corotation point $\omega_D=0$ (or $\xi_c=\omega/\omega_A$) is given by
\begin{equation}
s=\frac1{2({\omega'_i}^2\!-1)}\left[{\omega'_i}^2\!
-3\pm\sqrt{{\omega'_i}^4\!-14{\omega'_i}^2\!+17}\right].
\end{equation}
When ${\omega'_i}^2\geq7+4\sqrt{2}$, the eigenvalue is regular at the corotation point, since both indices are real and positive. When $0\leq{\omega'_i}^2<1$, indices are still real, and one of them is positive. 
This is the solution that is consistent with the matching condition at $\xi=0$ of the shooting method discussed in \S 3. When $1<{\omega'_i}^2\leq7+4\sqrt{2}$, the corotation point is singluar since two indices are real and negative in the region $1<{\omega'_i}^2\leq7-4\sqrt{2}$, and complex in the region $7-4\sqrt{2}<{\omega'_i}^2<7+4\sqrt{2}$. When ${\omega'_i}^2=1$, the eigenfunction has irregular singularity at the corotation point. Again, even though the eigenfunction is irregular at the corotation point, the physical eigenmodes on the real $\xi$-axis is regular.
Moreover, since all the coefficients of differential equation (\ref{eq2}) are real at the corotation point even when the eigenvalue is complex, the eigenfunction should be real at $\xi=\xi_c$.

Although irregular in the vicinity of the Alfv\'en singularities or the corotation point in most cases, eigenfunctions are not irregular in the physical sense unless the singularities are on the real axis. Instead, the oscillatory behavior and amplitude of the eigenfunction around the Alfv\'en singularities directly reflects the physical eigenfunction behavior on the real $\xi$-axis, especially if $\omega'_i$ is small, i.e., magnetic field is strong. 

When the eigenvalue $\omega$ and the index $s$ are both complex, the eigenfunction oscillates indefinitely in the vicinity of the singularity due to the imaginary component of the index $s$(see eq. [\ref{osc}]). In this case, the physical eigenfunction on the real $\xi$-axis also oscillates very rapidly in the vicinity of the point which is the projection of the complex singular point to the real $\xi$-axis, but the physical eigenfunction oscillates only finite times because the projected point is not a singular point.  

When the real component of $s$ is negative, the eigenfunction diverges at the singularity(eq. [\ref{osc}]). The amplitude of the physical eigenfunction on the real $\xi$-axis is large at the projected singular point on the real $\xi$-axis. However, since the projected point is not a singular point, the eigenfunction does not diverge at this point.
It is also clear that the eigenvalue is regular even if the singular point is a branch point, since we can choose the branch cut of the eigenvalue without crossing the real $\xi$-axis.

If the eigenvalue is real, the singularities are on the real $\xi$-axis and eigenfunction is irregular in the physical sense. We will discuss pure real eigenvalue cases in \S3. 

\section{Robustly Unstable Magnetoshearing Eigenmodes}
The eigenvalues of the wave equation (\ref{eq2}) are
calculated numerically by the shooting method with the boundary condition discussed in \S 2. Since the equation (\ref{eq2}) may have three singularities(corotation point and Alfv\'en resonances), we choose complex initial value to avoid Alfv\'en singularities on real axis, and integrate("shoot") the equation (\ref{eq2}) on the real $\xi$-axis from the left and right asymptotic boundaries(which are for removed from the any of particular singular behavior) to the corotation point, where their value and first derivative of the eigenmode are to be matched. If they are not matched, we change the eigenvalue appropriately until we can match. This iterative method is generally called the "shooting method" for eigenvalue problems. Spatial steps of the integration are smaller in the vicinity of the point where the Alfv\'en point is projected on the real $\xi$-axis than other regions, in case the eigenvalue is almost real but still complex. 
By using the Newton method to decrease 
errors of a trial function,
we obtain higher accuracy and faster convergence than the previous shooting codes.
This allows us to search for subtle singular and regular eigenfunctions over a wide range of parameter values.

In order to satisfy the boundary condition (\ref{bound}),
we impose $\bar{v}'_x/\bar{v}_x=k_\pm$ at the numerical
boundaries $\xi=\pm10$ (corresponding to the artificial infinity), where $k_\pm$ are the negative and positive
solutions of the quadratic equation given by inserting the
functional form $\bar{v}_x=\exp(k_\pm x)$ into the equation (\ref{eq2}). On the numerical boundaries, the leading term of the equation (\ref{eq1}) is the first term of the coefficient of $\bar{v}_x$, which is of the order of 1, and other terms are of the order of 0.01 or lower in our calculation. Our assumption for the boundary condition is justified as far as these estimations are valid.

\begin{figure*}[htb]
\begin{tabular}{cc}
\begin{minipage}{0.5\hsize}
\epsfxsize=3 truein \epsffile{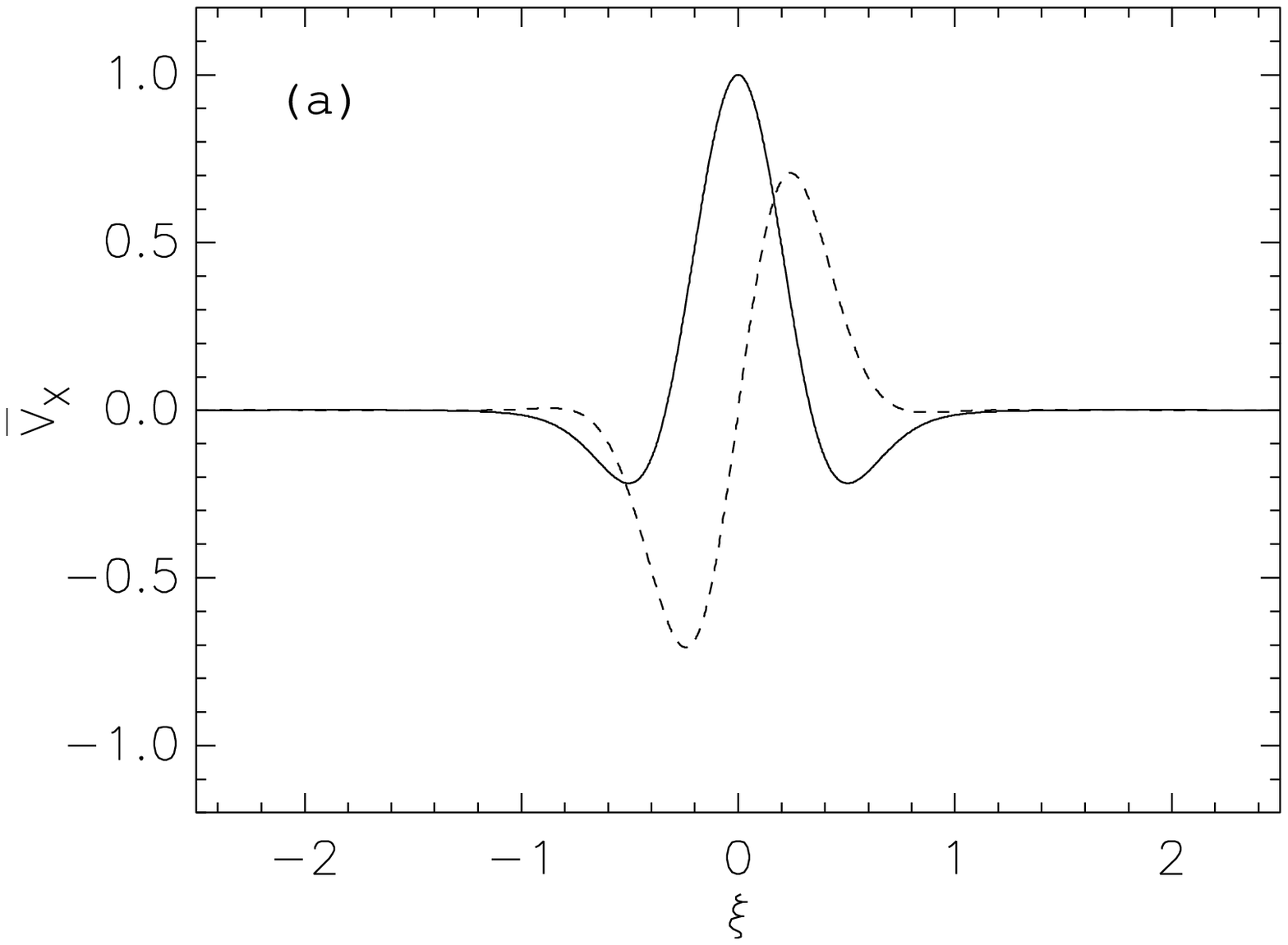}
\end{minipage}
\begin{minipage}{0.5\hsize}
\epsfxsize=3 truein \epsffile{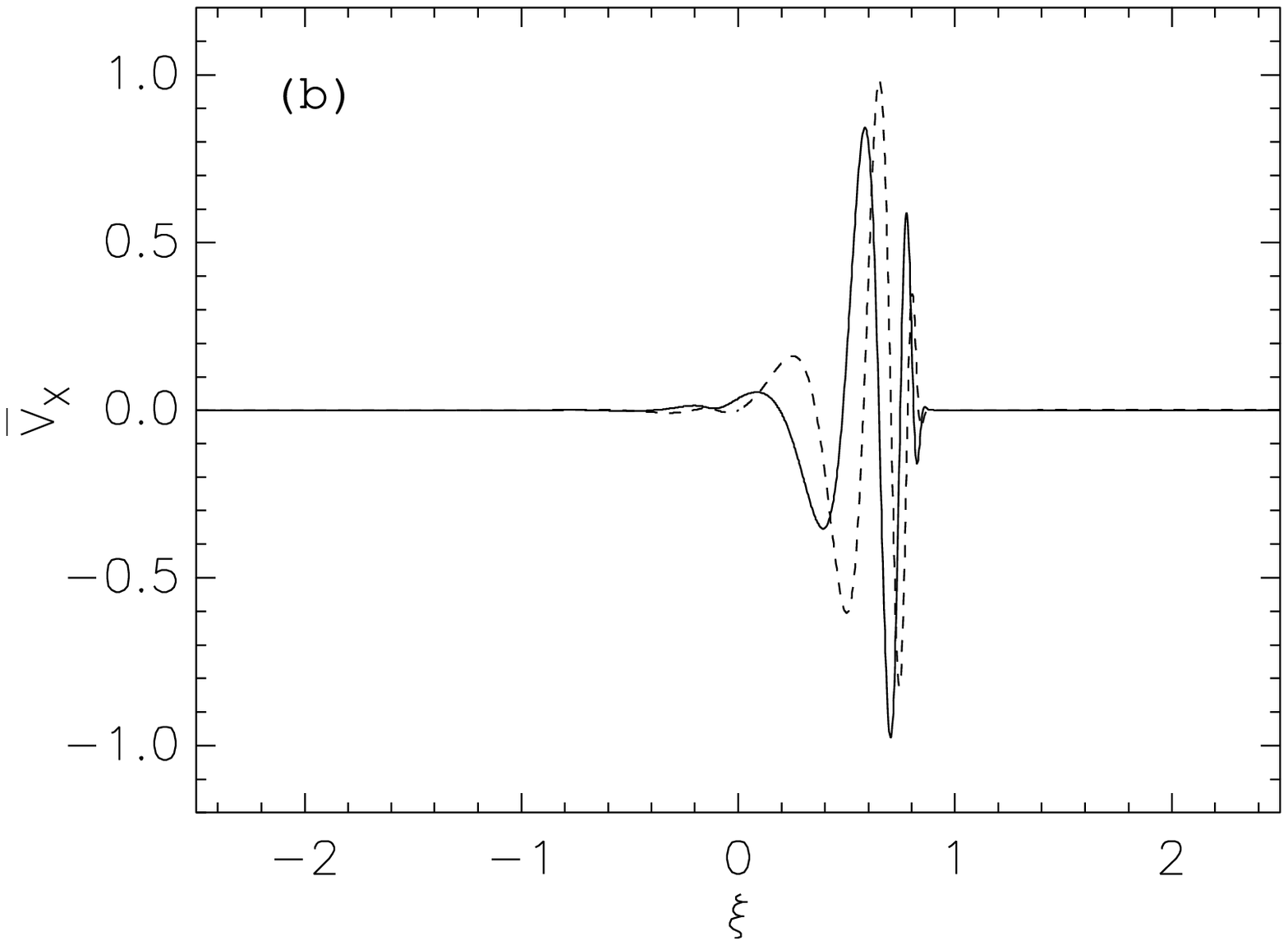}
\end{minipage}
\end{tabular}
\caption{Examples of the eigenfunctions of the nonaxisymmetric magnetoshearing instability in a Keplarian disk. {\it Solid curve,} real part of eigenfunction; {\it dashed curve,} imaginary part of eigenfunction. The model parameters are $\omega_A=0.1\Omega$ and $q=0.01$. ($a$) Fundamental pure imaginary
mode(eigenvalue is $\omega=0.00785\Omega\mbox{i}$);($b$) Complex mode(eigenvalue is $\omega=(0.00109+0.00039\mbox{i})\Omega$). The eigenfunctions are sandwiched by two Alfv\'en singularities $\xi=\pm1-\omega/\omega_A$.
}
\label{fig2}
\end{figure*}

Figure 2 shows examples of eigenfunctions $\bar{v}_x$ obtained by our shooting code when $\omega_A=0.1\Omega$
and $q=0.01$. The solid and dashed curves represent the real and imaginary
parts of the eigenfunction respectively. Figure 2a is for the fundamental pure imaginary
eigenvalue and Fig. 2b is for the complex eigenvalue. 
Since the eigenvalue of Fig. 2a
is pure imaginary, the real part of the eigenfunction is symmetric and
the imaginary part antisymmetric with respect to $\xi=0$, which is consistent with the
equation (\ref{sym1}). Figure 2b is the 
eigenfunction with a complex eigenvalue, which makes a pair with the
eigenvalue $-\omega^*$ whose eigenfunction is derived from the relation
(\ref{eq4}). These eigenfunctions are confined between two Alfv\'en 
singularities located at $\xi=\xi_{A\pm}\sim\pm1$, and they are real at $\xi=\xi_c$.

\begin{figure*}[htb]
\epsfxsize=3 truein \centerline{\epsffile{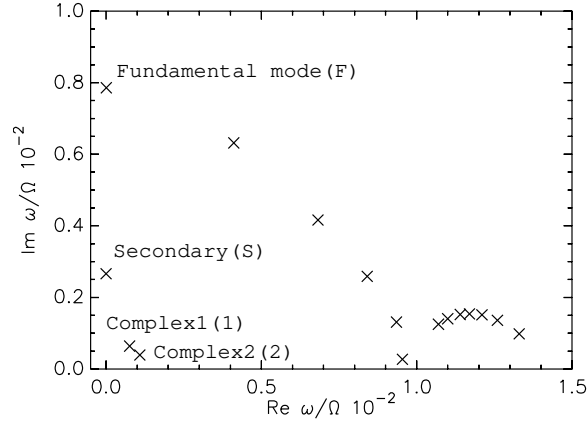}}
\caption{
Distribution of unstable eigenvalues of the magnetoshearing
instability in the upper complex $\omega$-plane when $\omega_A=0.1\Omega$ and
 $q=0.01$. Eigenvalues with Re($\omega)<0$ are shown, and all the complex 
eigenvalues have a paired unstable eigenvalue $-\omega^*$ in the region Re($\omega)<0$. There exist only two pure imaginary eigenmodes and many complex eigenmodes, which are not allowed to exist in self-adjoint system.
}
\label{fig3}
\end{figure*}

Figure 3 shows the distribution of eigenvalues in the upper complex
$\omega$-plane when $\omega_A=0.01$ and $q=0.01$. It shows only the eigenvalues
in the region Re($\omega)\geq0$, and all the complex eigenvalues have 
a paired eigenvalue $-\omega^*$ in the
region Re($\omega)<0$.
It is obvious that this non-self-adjoint system has complex eigenvalues, which never appear in a self-adjoint system (see Fig. 1).
There are only two pure imaginary eigenvalues,
which will be shown to merge by changing $\omega_A$ and $q$. 
We find that complex eigenvalues, which Matsumoto \& Tajima(1995) did not find, exist  
and have smaller imaginary part and grow slower in time than the fundamental eigenmode.

\begin{figure*}[htb]
\epsfxsize=3 truein \centerline{\epsffile{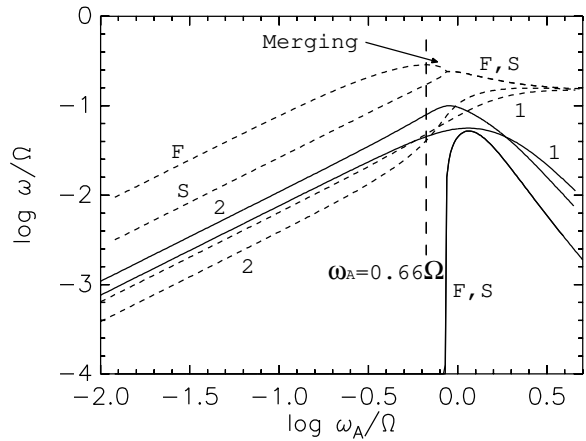}}
\caption{
The Alfv\'en frequency ($\omega_A=k_\parallel v_A$) dependence of eigenvalues of 
magnetoshearing instability when $q=0.01$. The dashed curves and solid curves show 
the growth rate Im($\omega$) and the real frequency Re($\omega$), respectively. The fundamental(F) and secondary(S) pure imaginary eigenmodes and two complex modes(1, 2) are shown, which correspond to the eigenmodes labeled F, S, 1 and 2 in Fig. 3, respectively, when $\omega_A/\Omega=0.01$.
Two pure imaginary
eigenvalues merge at $\omega_A\sim0.66\Omega$ to form complex eigenvalues. The
growth rate of all four eigenvalues saturates to $\omega\sim0.15\Omega\mbox{i}$
with increasing $\omega_A$.
}
\label{fig4}
\end{figure*}
Figure 4 shows the dependence of unstable eigenvalues on $\omega_A$ when $q=0.01$.
The solid(dashed) curves 
show the imaginary(real) part of the eigenvalues. When $\omega_A$ is small,
there exist two purely growing modes, which merge at $\omega_A\sim0.66\Omega$
and form complex eigenvalues. These modes were found in Matsumoto \& Tajima (1995) and the qualitative properties of this mode are about the same as found in Matsumoto \& Tajima. However, the merging point is slightly
greater than the earlier value and, more significantly, the growth rate does not decay significantly, even beyond $\omega_A=1.584\Omega$, where it was calculated to vanish in Matsumoto et al.. The fundamental mode acquires its maximum 
growth rate just before it merges with another(secondary) pure imaginary mode.
In addition, we find two new complex modes. These complex eigenmodes are also shown in Fig. 4, and the growth
rate of all these modes saturate to $\omega\sim0.15\Omega\mbox{i}$, the same saturation value for the fundamental mode with increasing $\omega_A$.

\begin{figure*}[htb]
\begin{tabular}{cc}
\begin{minipage}{0.5\hsize}
\epsfxsize=3 truein \epsffile{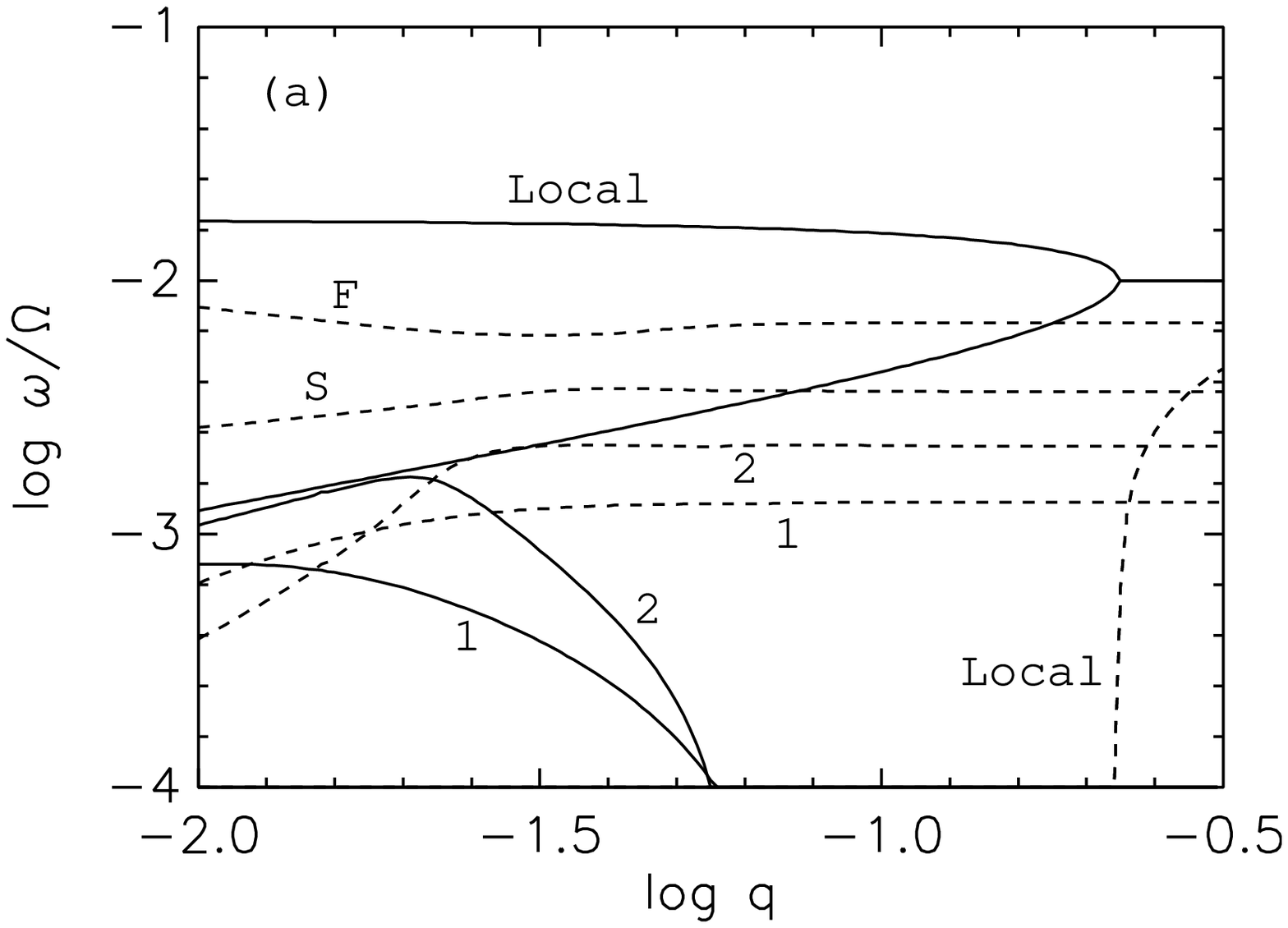}
\end{minipage}
\begin{minipage}{0.5\hsize}
\epsfxsize=3 truein \epsffile{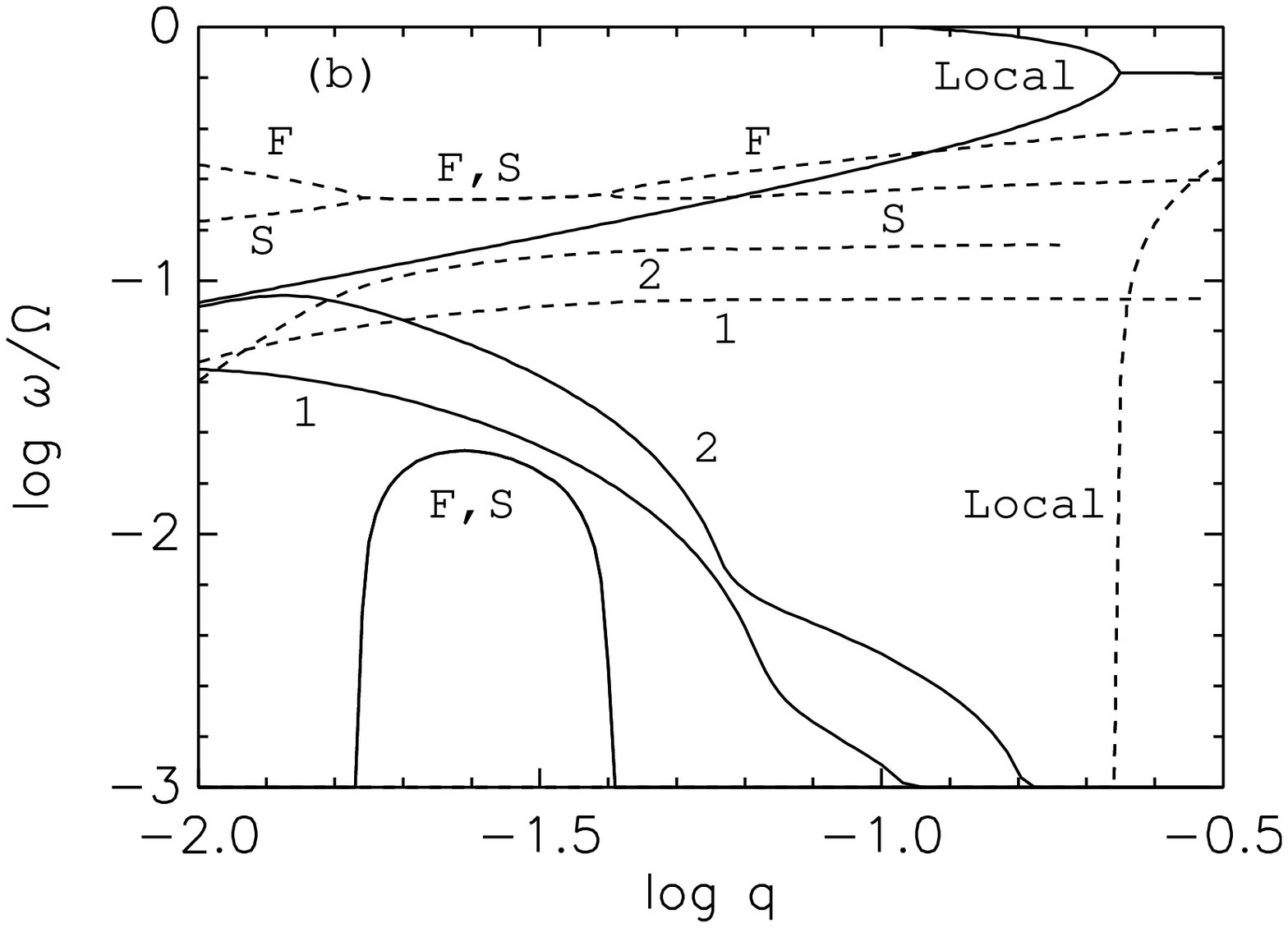}
\end{minipage}
\end{tabular}
\caption{
The $q$(=$k_y^2/k_z^2$) dependence of eigenvalues of magnetoshearing instability.
The dashed curves and solid curves show the growth rate Im($\omega$) and Re
($\omega$), respectively. The fundamental(F) and secondary(S) pure imaginary modes and two comlex modes(1, 2) are shown, which correspond to the eigenmodes labeled F, S, 1 and 2 in Fig. 4 when $q=0.01$. Eigenvalues calculated from local mode analysis are also shown.
(a) $\omega_A=0.01\Omega$. Two pure imaginary
eigenmodes(F, S) are always distinct, and the complex eigenmodes(1, 2) become pure imaginary,
whose growth rate saturates, with increasing $q$. (b) $\omega_A=0.66\Omega$. Two pure imaginary eigenvalues(F, S)
merge at $\log_{10}q\sim-1.8$ to form complex eigenvalue, and split again to become
imaginary at $\log_{10}q\sim-1.4$. The complex eigenmodes(1, 2) becomes pure imaginary with increasing $q$, and the growth rate saturates. In both cases, local modes are stable even in a region where nonlocal modes are unstable.
}
\label{fig5}
\end{figure*}

Figure 5 shows the dependence of eigenvalues on $q$ when 
$\omega_A=0.01\Omega$ (Fig. 5a) and $0.66\Omega$ (Fig. 5b).
The fundamental(F) and secondary(S) pure imaginary modes and two complex modes(1, 2) are shown in Fig. 5, which correspond to the eigenmodes in Fig. 4, when $q=0.01$. 
Eigenvalues calculated from local analysis are also shown in Fig. 5, which we discuss later.  
Two pure imaginary modes are always
distinct when $\omega_A=0.01\Omega$ (two upper modes in Fig. 5a).
However, when $\omega_A=0.66\Omega$, these two modes merge and become complex
at $\log_{10}q\sim-1.8$ and split to become pure imaginary again at 
$\log_{10}q\sim1.4$. 
The eigenvalues of the other two complex modes become pure imaginary when $q$
exceeds a certain value ($\log q=-1.25$ for $\omega_A=0.01\Omega$, $\log q=-0.75$ for $\omega_A=0.66\Omega$), and the growth rate for those modes saturates
with increasing $q$.

Next, we compare the nonlocal eigenfunction results with the local(Fourier) dispersion
relation. By replacing $d/dx$ in equation
(\ref{eq1}) with a constant $\mbox{i}k_x$ around $x=0$ and assuming that the unperturbed magnetic
field is toroidal($B_x=B_z=0$), the local solution in the regime
\[
|\omega|\sim\omega_A\ll\omega_I\equiv\sqrt{\Omega^2k_z^2/(k_x^2+k_y^2+k_z^2)}
\]
is (Matsumoto \& Tajima 1995)
\begin{equation}
\omega^2=\frac3{2}\left[1-\frac3{2}q\pm\frac3{2}\sqrt{(q-2)(q-\frac2{9})}\right]\omega_A^2.\label{eq25}
\end{equation}
This local dispersion relation (\ref{eq25}) shows that pure real eigenmodes appear in the region $q<\frac{2}{9}$, pure imaginary eigenmodes in $q>2$, and complex in $\frac{2}{9}<q<2$. 
However, when $q$ is small, i.e., the perturbation is almost parallel to the magnetic field, nonlocal eigenmodes are unstable in both $\omega_A=0.01\Omega$ and $0.66\Omega$ (see Figs. 5a, b), 
and the growth rate of each mode does not have strong dependence on $q$.
We conclude that replacing $\partial/\partial x$ by a single wavenumber
 $k_x$ is invalid for these modes since such eigenmodes oscillate 
very rapidly in the vicinity of the Alfv\'en points in a pronounced fashion(see Fig. 2b). In other words, the spatial variation of the wavenumber in the radial direction is essential for the modal analysis of the magnetoshear instability. Radial dependency of the wavelength also prevents us from applying the WKB method to this model. The WKB method requires the wavelength of the eigenmodes $L_e$ is smaller than the shear scale length $L_s(L_e/L_s\ll1)$, which may be satisfied around the corotational and Alfv\'en singularities, but the wavelength is comparable to the shear scale length in other regions$(L_e/L_s\simeq 1)$.

To show that these eigenmodes are discrete, let us first show the existence
of the Alfv\'en continuum on the real $\omega$-axis.
The wave equation (\ref{eq2}) has a
solution for any real $\omega$ for which $\omega_D^2=\omega_A^2$ for some
$x$. It follows that the spectrum of this mode 
is continuous, and the Alfv\'en continuum extends to the entire real $\omega$
by choosing some $\bvec{B}, k_y$ and $k_z$, which is different from the model chosen by Ogilvie \& Pringle (1996) in which the Alfv\'en continuum is restricted by the boundary condition. The eigenmodes with the pure imaginary eigenvalue that we
have shown are obviously not in this class.

When $\bvec{B}=0$, the Kelvin-Helmholtz modes can be derived, which are stable in accretion disks. In this limit,
equation (\ref{eq1}) reduces to
\begin{equation}
\frac{d^2\bar{v}_x}{d{}x^2}+\left[-k_\parallel^2+\frac{\Omega^2k_z^2}{\omega_D^2}\right]\bar{v}_x=0,
\end{equation}
and when $k_z=0$, it has a simple solution $\bar{v}_x=\exp[-k_y|x|]$, 
which satisfies the boundary condition (\ref{bound}). The first
derivative of this class of solutions is discontinuous at $x=0$, which vanishes
with introducing dissipation. The Kelvin-Helmholtz instability also
has continuous eigenvalues,
but this class of solutions is eliminated in our calculation because of
the matching condition of the shooting method, which requires the eigenfunction
and its first derivative to be continuous.
We search for eigenvalues by choosing initial values in the region
$0<\omega_r/\omega_A<1$ and $0<\omega_i/\omega_A<1$.
We find that of the initial values converge to one of
the eigenvalues in Fig. 3 and we conclude that
all of the eigenmodes are discrete.

\begin{figure*}[htb]
\epsfxsize=3 truein \centerline{\epsffile{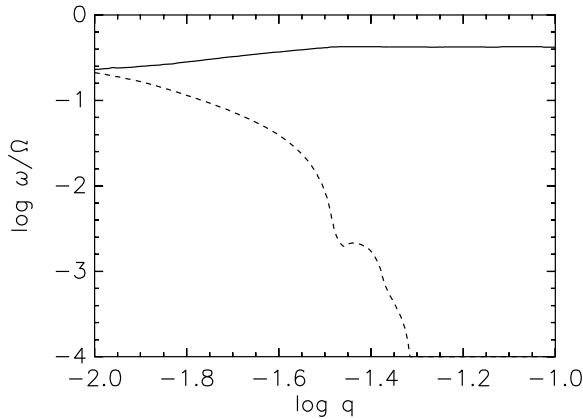}}
\caption{
The example of a complex eigenmode that becomes pure real
value with increasing $q$ when $\omega_A=0.66\Omega$. The dashed curve and solid curve show the growth rate Im($\omega$) and the real frequency Re($\omega$), respectively. 
The eigenvalue is complex when $q$ is small, which becomes real with increasing $\omega_A$.
}
\label{fig6}
\end{figure*}
\begin{figure*}[htb]
\epsfxsize=3 truein \centerline{\epsffile{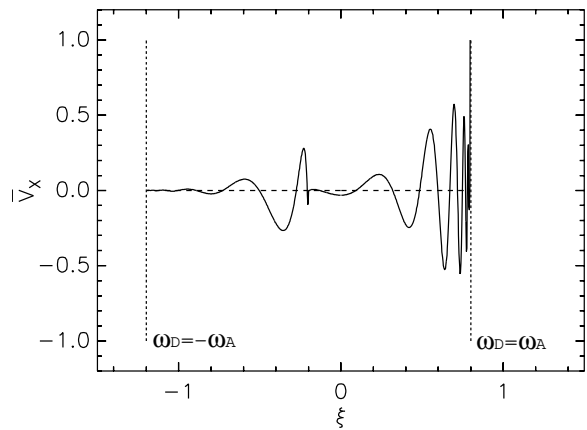}}
\caption{
The mode with a pure real frequency that has the Alfv\'en singularities at $\omega=\pm\omega_A$, where the mode energy evaneces. The eigenfunction oscillates indefinitely towards $\omega=\omega_A$, and also towards $\omega=-\omega_A$ with small amplitude.
}
\label{fig7}
\end{figure*}

In Fig. 6 we show another eigenmode whose eigenvalue gradually
becomes real with increasing $q$, when $\omega_A=0.66\Omega$. However, when the eigenvalue is real,
we have already shown that the index of the eigenfunction $s$ is pure imaginary
in the vicinity of the Alfv\'en singularities (Eq. (\ref{real})) and that the eigenfunction has the form
\begin{equation}
\bar{v}_x=\exp\left[\mbox{i}|s|\log(\xi-\xi_{A\pm})\right],
\end{equation}
from equation (\ref{osc}).
Such eigenfunctions oscillate indefinitely toward the singularity and the
function can take any value between $-1<\bar{v}_x<1$. This indicates that
the function in the inner region $\xi_{A-}<\xi<\xi_{A+}$ and outer region
$\xi<\xi_{A-},\xi<\xi_{A+}$ is discontinuous at the Alfv\'en singularities.
Thus the boundary condition for the continua cannot be that of shooting from the outside $|\xi|=\infty$ toward the inside, but we should shoot from inside toward the singularities.

Figure 7 shows an example of eigenfunction in the inner region when $\omega_A=0.01\Omega$, $\omega=0.002\Omega$ and $q=0.01$(see eq. (\ref{real}) and arguments for detail of this mode). 
However, the eigenmode in Fig. 6 is continuous even at the Alfv\'en
singular points, since the integration by a finite spatial step brings in an effective dissipation, which 
is not the case for the pure real eigenvalue. Instead, the eigenmode 
becomes continuous because of the numerical dissipation. Although this numerical eigenfunction is different from the theoretical eigenfunction beyond the passage of the singularity, the fact of continua remains the same for two different reasons.
It should be pointed out that in real physical situation there always exists a dissipation even for a nearly ideal MHD system. The dissipation prevents the eigenmode from blowing up on the Alfv\'en singular points, keeping the energy of the eigenmode finite. The numerical dissipation affects this eigenmode in the same manner mathematically as the real dissipation does, by passing oscillations through the singularity barrier and avaraging oscillation in a finite spatial step. Thus the numerically obtained eigenmode, though different from theoretically expected continua, may be regarded as realistic. 

\begin{figure*}[htb]
\begin{tabular}{cc}
\begin{minipage}{0.5\hsize}
\epsfxsize=3 truein \epsffile{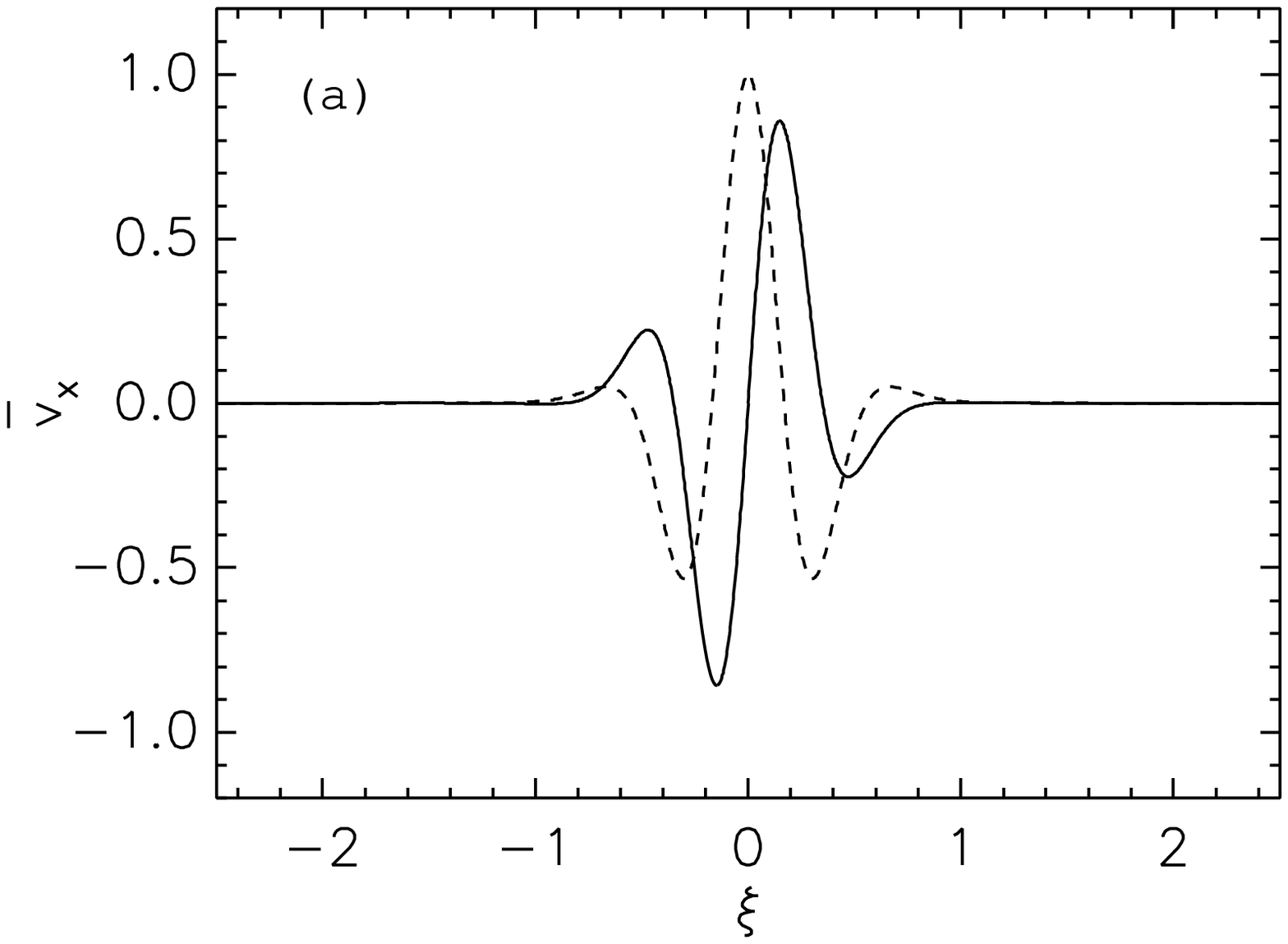}
\end{minipage}
\begin{minipage}{0.5\hsize}
\epsfxsize=3 truein \epsffile{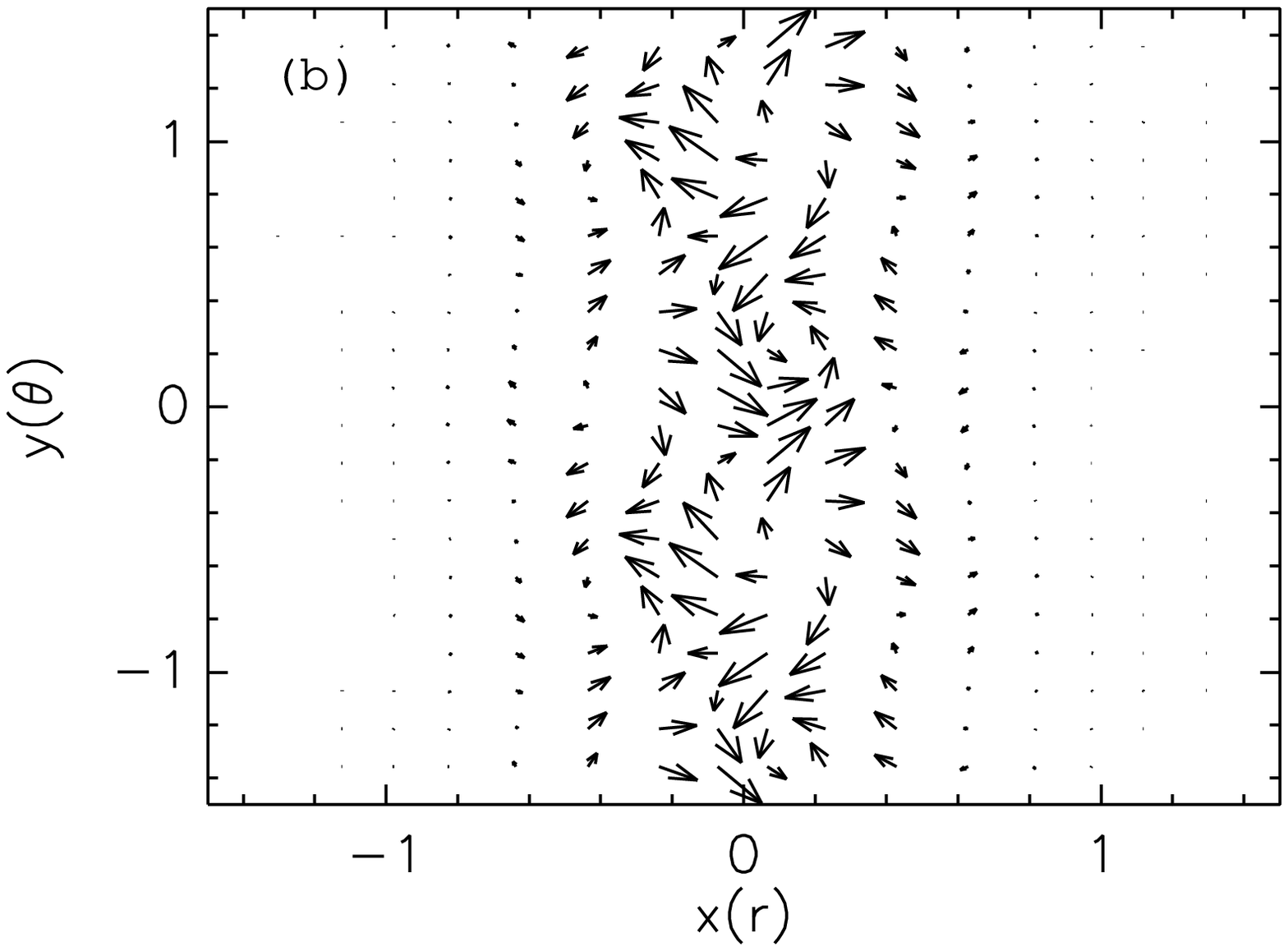}
\end{minipage}
\end{tabular}
\caption{
An example of $\bar{v}_y$ calculated from the fundamental pure
imaginary mode (Fig. 2a) and the velocity field in $xy$ plane by the fundamental mode. The solid curve and dashed curves show the real and imaginary part of $\bar{v}_y$, respectively. (a) $\bar{v}_y$ is almost out of phase with respect to $\bar{v}_x$. (b) Velocity field is created in $xy$-plane by the fundamental mode. Vortices are created between two Alfv\'en singularities, and they will overlap with other eigenmodes excited at various $x$-positions to expand the unstable region.
}
\label{fig8}
\end{figure*}

Finally, we describe the physical behavior of the eigenmodes in accretion disks.
The expression of $\bar{v}_y$ in terms of $\bar{v}_x$ is derived from the
continuity equation(\ref{cont}),
\begin{eqnarray}
\lefteqn{\bar{v}_y=\frac{\mbox{i}}{1+q}\Bigg[\frac{q}{k_z}\frac{\partial}{\partial\xi}}\nonumber\\
&&\left.-\frac{\omega_D\Omega}{2(\omega_D^2-\omega_A^2)}\left(3\frac{\omega_A^2}{\omega_D^2}+1\right)\right]
\bar{v}_x.
\end{eqnarray}
Figure 8a shows $\bar{v}_y$ calculated from the fundamental pure
imaginary mode (Fig. 2a) and the velocity field created in the $xy$-plane by the fundamental eigenmode is shown in Fig. 8b. 
The eigenfunction of $\bar{v}_y$ is also trapped between two Alfv\'en singularities. Since both $\bar{v}_x$ and $\bar{v}_y$ are almost out of phase with each other, the velocity field created by the fundamental eigenmode consists of vortices in $xy$-plane which are the seeds of nonlinear instability(\cite{mat95}).
Note that since we use the frame rotating with angular velocity $\Omega$, there is no unique origin $x$ in the $xy$-plane. Thus such unstable eigenmodes excited at various $x$-positions will overlap with each other to expand the unstable region in the $x$-direction.

\section{Summary}
We have shown the existence of the discrete unstable
nonaxisymmetric magnetoshearing instability eigenmodes. 
Since we assumed the exponentially decaying boundary condition 
(Eq. [\ref{bound}]) for the radial component of velocity,
our modes are independent of the effect of the boundary condition.
In our accretion disk theory, this robust instability occurs without any unrealistic
disk edge boundary condition in infinite linear shear flow.
The scalelength of a single eigenmode in the $x$ direction $\Delta x$ is
determined by the local strength of magnetic field, 
the direction and amplitude of the wavenumber and the magnitude
of the angular velocity (see Eqs. [\ref{omd}] and [\ref{oma}]),
since the eigenfunction is buffeted by the Alfv\'en singular
points $\omega_D=\pm\omega_A$. If the magnetic field is pure toroidal, $\Delta x=2\omega_A/3\Omega k_y=2v_A/3\Omega\simeq2/3(v_A/C_s)H$, where $C_s$ is the sound speed and $H=C_s/\Omega$ is the thickness of the disk. When $v_A\ll C_s$, the mode is localized in the radial direction with the scalelength smaller than the thickness of the disk. If the magnetic field has azimuthal component, $\Delta x$ is propotional to $k_\parallel/k_y$(\cite{mat95}). 
In both cases, our infinite boundary condition is sufficient if the scalelength of the eigenmode is smaller than $H$. The curvature of the magnetic field is also small if $\Delta x\ll H$. However, for nearly axisymmetric perturbations($k_y\ll k_\parallel$), the eigenmode have a large radial scalelength, and the infinite boundary condition may not valid. Density gradient and geometrical effects become important in this case.

Our result of the growth rate of unstable modes agrees with that of
Matsumoto \& Tajima (1995) in the region $\omega_A\leq\Omega$. We
have found complex eigenvalues with smaller growth rates than
the fundamental pure imaginary eigenmode. When $\omega_A$ is
larger than $\Omega$, two pure imaginary eigenmodes merge,
the results of which is the same as found in Matsumoto \& Tajima. However, our
result shows that the growth rate saturates with increasing
$\omega_A$. This indicates that the accretion disk is unstable
even if the Alfv\'en frequency is comparable to the angular velocity,
a case of strong magnetic fields.

The comparison of the nonlocal and local dispersion relations 
demonstrates where and how the local Fourier mode approximation fails to be accurate
for this nonaxisymmetric instability. The wavenumber dependence
of the eigenvalue shows that the nonlocal modes are unstable 
even in the region $q=k_y^2/k_z^2\ll1$, where the local dispersion
relation has only a stable solution. Furthermore, we have found that two
pure imaginary eigenvalues merge to be complex and split
into two pure imaginary again with increasing $q$ in a region
where the solutions of the local dispersion relation are purely real.  
Overall, as Fig. 5 indicates, the discrepancy of the local theory from the correct
nonlocal theory amounts to not just a quantitative level but a qualitative deviation.

We have developed the mathematical and physical theory of a system of nonself-adjoint
differential equations for the first time. In astrophysics, nonself-adjointness always 
appears whenever there is a shear flow, which is mathematically unsolved so far.
We find the relationship of complex
eigenvalues, which a self-adjoint system does not have,
by a general approach to the nonself-adjoint system. 
A pair of eigenvalues $\omega$ and $-\omega^*$ relate to each other
in our model, since our model is symmetric with respect to the $x$-axis.
Even if there is no such symmetry, we conclude that four eigenmodes
make up a group in the nonself-adjoint system in general.

Although our model of the nonaxisymmetric mode has been
linear in Cartesian coordinates and ignored the effect of diffusion, 
analysis in Section 3
suggests how the eigenmode grows to enter a nonlinear stage, and
how it explains momentum transport in accretion disks.
Since there is no particular sense in the radial direction in Cartesian coordinates, 
there exists no specific calibration of momentum transport in the linear stage.
The eigenmodes can be excited, however, between any pair of 
Alfv\'en singularities in the radial direction (the $x$-axis) and create vortices in the disk plane (the $xy$-plane),
as shown in Fig. 8b. The eigenmode with the fundamental
eigenvalue dominates in time for a given radial co-rotation point.
For another (arbitrary) co-rotation point, the same applies.
These eigenmodes can overlap with each other
to form greater vortices. In this stage, the nonlinear effect gives rise to
anomalous magnetic viscosity that underlies
the momentum transport needed to explain astrophysical disks.
Matsumoto \& Tajima demosntrated non-linear evolution of the
eigenmode by three-dimentional MHD simulation with the
shearing-box model. The overlap of eigenmodes excited at various $x$s' was shown. They also calculated the magnetic viscosity parameter
\begin{equation}
\alpha_B=-\frac{\langle\delta B_x\delta B_y\rangle}{4\pi\rho C_s^2}<\frac{\langle\delta B^2\rangle}{4\pi\rho C_s^2}\simeq\frac{\omega_A^2}{k_\parallel^2C_s^2}<\frac{1}{(k_\parallel H)^2}
\end{equation}
where the notation $\langle\delta B_x\delta B_y\rangle$ etc., denotes the spatial avarage.
They found that when the poloidal field is dominant, the magnetic viscosity is $\alpha_B\sim O(0.1)$, which corresponds to $\alpha$ in dwarf novae during the bursting phase.

We conclude that the results of Matsumoto \& Tajima are correct
and that the robust mode in magnetized accretion disks
is of the magnetoshearing origin. This mode should be
dominant in nonlinear theory and our linear analysis 
supports the results from the three-dimensional
simulation in Matsumoto \& Tajima, which explained anomalous momentum
transport in accretion disks. 

\acknowledgements

The work was supported by the US Department of Energy and NSF ATM 98-15809.
TT is currently on leave at LLNL.

\appendix
\section{Exisitence of the Localized Growing Mode in the Infinite Shearing Flow}
Ogilvie \& Pringle's argument of the non-existence of localized growing modes
(\cite{ogi96}, Appendix C), is incorrect for the following reasons.

First, we rewrite the integrated wave equation (\cite{ogi96}, [C8]) without introducing a parameter $\lambda$ for simplicity,
\begin{equation}
\int^{+\infty}_{-\infty}\left[\left|\frac{d{}y}{d{}x}\right|^2+(k^2-f)|y|^2
\right]dx=0,\label{C8}
\end{equation}
where $k^2=k_y^2+k_z^2$ and the eigenfunction
$y(x)$ is assumed to decay exponentially as $|x|\to\infty$, as we assume for the boundary condition. They also assume a uniform magnetic field, which makes equation (\ref{C8}) qualitatively equivalent to our wave equation (\ref{eq2}).
The function $f(x)$ is given by
\begin{eqnarray}
\lefteqn{f(x)=\frac{(k_yu')^2\omega_A^2}{[(i\omega_i-k_y u'x)^2-\omega_A^2]^2}+k_z^2\left[\frac{2\Omega u'}{[(i\omega_i-k_y u'x)^2-\omega_A^2]}\right.}\nonumber\\
&&+\left.\frac{4\Omega^2(i\omega_i-k_y u'x)^2}{[(i\omega_i-k_y u'x)^2-\omega_A^2]}\right],
\end{eqnarray}
where $u(x)$ is the shearing velocity, and $f(x)$ has two Alfv\'en singular points in the lower half-plane.
We show here that equation (\ref{C8}) can be satisfied 
because of those two Alfv\'en singular points, which
makes the integral (\ref{C8}) equal to zero if the complex frequency is 
 properly chosen. 
 The positions of the Alfv\'en singular points are given by
\begin{equation}
x=\frac{\mbox{i}\omega_i\pm\omega_A}{k_y u'}
\end{equation}
where $\omega_i$ is the imaginary part of the eigenfrequency, and $k_y$ is the wavenumber parallel to the shear flow, respectively.
Note that the total integrand of the equation (\ref{C8}) becomes positive and decays to zero as $|x|\to\infty$ since 
$f(x)$ decays to zero as $|x|\to\infty$.
 
Now, if $k^2>|f(x)|$ for any $x$, the integrand is always positive and equation ({\ref{C8}) can not be satisfied with any eigenfrequency.
However, if the singularity is near enough to or on the real axis, $|f(x)|$ becomes large enough to satisfy $|f(x)|>k^2$ in the vicinity of the singularities and the total integral can take any value. Note that the real part of the eigenfrequency does not affect the position of the singularity. 
Furthermore, if the eigenvalue is complex, as we calculated in \S 3, the integral can equal to zero without crossing any branch cut of the singularities.  
Therefore, localized growing modes should exist by choosing eigenfrequencies which 
makes the integral of the equation (\ref{C8}) equal to zero.
In other words, the contour of the integral (\ref{C8}) is not free to pass around the circle in the upper half plane, since the complex conjugate of the eigenfunction $y^*$ has two singularities in the upper half plane with branch cuts, which any half circle in the upper half plane with sufficient large radius should cross. 
  
Thus the statement "The integral relation (C8) therefore cannnot be satisfied" in line 25 of p.164 of Ogilvie \& Pringle was incorrect, and our calculation is valid.

\end{document}